\documentclass[referee]{mn2e}
\usepackage{epsfig}
\usepackage{txfonts}

\title[Thermal evolution of rotating hybrid stars]{Thermal evolution of rotating hybrid stars}

\author[Miao Kang , Xiao-Ping Zheng ]{Miao Kang $^{}$\thanks{E-mail:kangmiao@phy.ccnu.edu.cn}, Xiao-Ping Zheng
$^{}$\thanks{E-mail:zhxp@phy.ccnu.edu.cn}
\\ The Institute of Astrophysics, Huazhong Normal
University, Wuhan 430079, China}

\date{Accepted 0000, Received 0000}
\pagerange{\pageref{firstpage}--\pageref{lastpage}} \pubyear{0000}

\begin{document}
\label{firstpage} \maketitle

\begin{abstract}
As a neutron star spins down, the nuclear matter continuously is
converted into quark matter due to the core density increase and
then latent heat is released. We have investigated the thermal
evolution of neutron stars undergoing such deconfinement phase
transition. We have taken into account the conversion in the frame
of the general theory of relativity. The released energy has been
estimated as a function of change rate of deconfinement baryon
number. Numerical solutions to cooling equation are obtained to be
very different from the without heating effect. The results show
that neutron stars may be heated to higher temperature which is
well-matched with pulsar's data despite onset of fast cooling in
neutron stars with quark matter core. It is also found that
heating effect has magnetic field strength dependence. This
feature could be particularly interesting for high temperature of
low-field millisecond pulsar at late stage. The high temperature
could fit the observed temperature for PSR J0437-4715.
\end{abstract}

\begin{keywords}
stars: neutron  --- dense matter --- stars: rotation
\end{keywords}

\section{Introduction}
 The composition of Neutron star (NS) interior is still poorly
  known due to the uncertainties of nuclear physics. NS cooling is an important tool for the study of
  dense matter. By comparing
  cooling models with thermal emission data from observations, we
  can gain insight into the equation of state (EOS) of dense matter
  inside NSs. After a supernova explosion a newly formed NS first
  cools via various neutrino emission mechanisms before the
  surface photon radiation take over. Only slower neutrino cooling
  mechanisms such as the modified Urca, plasmon neutrino and
  bremsstrahlung processes,  should occur when the interior
  density is not high. However, for higher core density the more 'exotic' extremely fast
  cooling  processes take over. Cooling of hybrid stars (NSs with quark matter core)
is just one of those examples.

 However, heating effect on the cooling of compact stars is an
 important factor. Several heating mechanisms have been
 extensively discussed, for example, the dissipation of
 rotational  energy due to viscous damping (Zheng \& Yu \cite{2}) and mutual friction
 between superfluid and normal components of the star(Shibazaki $\&$ Lamb \cite{3}) and release
 of strain energy stored by the solid crust due to spin-down
 deformation(Cheng et al. \cite{4}).
 These heating processes are closely related with the rotation
 evolution of a star. It is well-known that a neutron star spin-down due to magnetic dipole radiation.
  Because of spin-down
 compression, the interior density of the star will gradually
 increase. For hybrid stars, this results in, little by little, the
 transformation of hadron matter into quark matter in the
 interior. It will lead to the release
 of latent heat if the transition is the first-order one. The generation of the energy increases internal
 energy of the star. It will be called deconfinement heating (DH).
 Deconfinement process has ever been investigated in strange
 stars, where neutron drops at bottom of a crust drip on quark
 matter surface to be instantaneously dissolved into quark matter
 (Haensel $\&$ Zdunik \cite{7}, Yuan $\&$ Zhang \cite{8}, Yu \& Zheng \cite{9}).
 What we will here deliberate is slow transition process: nuclear
 matter undergoing from hadron matter phase (HP) to mixed
 hadron-quark matter phase (MP) and then to quark matter phase
 (QP) with gradually increasing density.

 From the point of view
 above, the change of the internal structure of the compact star
 due to rotation has to be evaluated within general relativity
 theory. Using the method of perturbation theory(Hartle \cite{10}, Chubarian et al \cite{17}), we are going to
 investigate evolution of phase transition region during the
 spin-down evolution of the rotating star. And then we also
 calculate changes in confined baryon numbers and quark numbers.
 The heat luminosity can be estimated with proportion to the
 change rates.

 Of course, the star is not quite in weak interaction equilibrium state during
spin down. The departure from the chemical equilibrium indeed lead
to the rotochemical heating in a rotating neutron star
(Reisenegger \cite{5}, \cite{6}), as well as a hybrid star. The
direct Urca processes would be triggered and dominates the
rotochemical heating in our models. We find, however, the effect
of rotochemical heating is much smaller than the deconfinement
heating by estimating the both net heating rates for the sequences
of different parameters. Hence, we don't consider the effect of
rotochemical heating in following study.

In this work, we take Glendenning's hybrid stars model
(Glendenning \cite{11}). The hadronic matter equations of
state(EOS) in the framework of the relativistic mean field theory
and MIT bag model of quark matter are used to construct the model
of hybrid stars, but medium effect of quark matter has been
considered in quasiparticle description(Schertle et al \cite{12}).
We
 choose the simplest possible nuclear matter composition, namely
neutrons, protons, electrons, and muons (npe$\mu$ matter) and
ignore superfluidity and superconductivity.

 The plan of this paper is as follows. In Sec. 2 we introduce
 rotating
hybrid star model. The DH effect is considered in Sec. 3. The
cooling curves and the corresponding explanations are presented in
Sec. 4. The conclusion and discussions are summarized in Sec. 5.

\section*{2. Rotating hybrid stars model}

Quark deconfiment phase transition is expected to occur in neutron
matter at densities above the nuclear saturation density
$n_{b}=0.16 fm^{-3}$. Since many theoretical calculations have
suggested that deconfinement transition should be of first order
in low-temperature and high-density area (Pisalski $\&$ Wilczek
\cite{13} and Gavai et al. \cite{14}), one may expect a MP during
the transition. Most of the approaches to deconfinement matter in
neutron star matter use a standard two-phase description of EOS
where the HP and the QP are modelled separately and resulting EOS
of the MP is obtained by imposing Gibbs conditions for phase
equilibrium with the constraint that baryon number as well as
electric charge of the system are conserved (Glendenning
\cite{15}, \cite{11}).

The Gibbs condition for mechanical and chemical equilibrium at
zero temperature between the HP and the QP reads
\begin{equation}
p_{HP}(\mu_{n},\mu_{e})=p_{QP}(\mu_{n},\mu_{e}).
\end{equation}
where $p_{HP}$ is pressure of HP and $p_{QP}$ is the pressure of
QP. We use the EOS of  the relativistic mean field model
(Glendenning \cite{11}) for hadron matter and employ a effective
mass bag-model EOS for quark matter(Schertle et al. \cite{12}).
Only two independent chemical potentials remain according to the
corresponding two conserved charges of the $\beta$-equilibrium
system. The total baryon number $N_{B}$ as well as electrical
charge Q
\begin{equation}
n_{B}=\frac{N_{B}}{V}=\chi n_{QP}+(1-\chi) n_{HP}.
\end{equation}
\begin{equation}
0=\frac{Q}{V}=\chi q_{QP}+(1-\chi) q_{HP}.
\end{equation}
where $\chi=V_{Q}/V$ is the volume fraction of quark matter in the
MP. Taking the charge neutral EOS of the HP, Eq.(1), (2) and (3)
for MP and the charge neutral EOS of the QP, we can construct the
full hybrid star EOS. In Fig.1 we show the model EOS with
deconfinement transition which is the typical scheme of a first
order transition at finite density with MP. We choose the
parameters for hadronic matter EOS which have given by Glendenning
\cite{11} and quark matter EOS with s quark mass $m_{s}=150$MeV,
bag constant $B^{1/4}=160$MeV, coupling constant $g=3$.

With the evaluated hybrid star EOS presented above we now turn to
analyse the structure of the corresponding rotating hybrid stars.
Using the Hartle's perturbation theory \cite{10}, Chubarian et al
\cite{17} have studied the change of the internal structure of the
hybrid stars due to rotation.
 In this paper, we
also apply Hartle's approach to investigate the structure of
rotating hybrid stars. Hartle's formalism is based on treating a
rotating star as a perturbation on a non-rotating star, expanding
the metric of an axially symmetric rotating star in even powers of
the angular velocity $\Omega$. The metric of a slowly rotating
star to second order in the angular velocity $\Omega$, can be
written as
\begin{eqnarray}
ds^{2}=-e^{\nu(r)}[1+2(h_{0}+h_{2}P_{2})]dt^{2}+
e^{\lambda(r)}[1+\frac{2(m_{0}+m_{2}P_{2})}{(r-2M(r))}]dr^{2} \nonumber\\
+r^{2}[1+2(v_{2}-h_{2})P_{2}]
\{d\theta^{2}+\sin^{2}\theta[d\phi-w(r,\theta)dt]^{2}\}+O(\Omega^{3})
\end{eqnarray}
Here $e^{\nu(r)}$,$e^{\lambda(r)}$ and $M(r)$ are functions of $r$
and describe the non-rotating star solution of the
Tolman-Oppenheimer-Volkov (TOV) equations.  $P_{2}=P_{2}(\theta)$
is the $l=2$  Legendre polynomials. $\omega$ is the angular
velocity of the local inertial frame and is proportional to the
star's angular velocity $\Omega$, whereas the perturbation
functions $h_{0},h_{2},m_{0},m_{2},v_{2}$ are proportional to
$\Omega^{2}$. we assume that matter in the star is described by a
perfect fluid with energy momentum tensor
\begin{equation}
T^{\mu\nu}=(\epsilon+P)u^{\mu}u^{\nu}+Pg^{\mu\nu}
\end{equation}
The energy density and pressure of the fluid are affected by the
rotation because the rotation deforms the star. In the interior of
the star at given $(r,\theta)$, in a reference frame that is
momentarily moving with the fluid, the pressure and energy density
variation is respectively
\begin{equation}
\delta P(r,\theta)=[\epsilon(r)+P(r)][ p_{0}^{*}+
p_{2}^{*}P_{2}(\theta)]
\end{equation}
\begin{equation}
\delta
\epsilon(r,\theta)=\frac{d\epsilon}{dP}[\epsilon(r)+P(r)][p_{0}^{*}+
p_{2}^{*}P_{2}(\theta)]
\end{equation}
here, $p_{0}^{*}$ and $p_{2}^{*}$ are dimensionless functions of
$r$, proportional to $\Omega^{2}$, which describe the pressure
perturbation. The rotational perturbations of the star's structure
are described by the functions
$h_{0},m_{0},p_{0}^{*},h_{2},m_{2},v_{2},p_{2}^{*}$. These
functions are calculated from Einstein's field equations. The
effect of rotation described by the metric on the shape of the
star can be divided into contributions: A spherical expansion
which changes the radius of the star, and is described by the
functions $h_{0}$ and $m_{0}$. The other part is a quadrupole
deformation, described by functions $h_{2}$, $v_{2}$ and $m_{2}$.
As a consequence of these contributions, the difference between
the gravitational mass of the rotating star and the non-rotating
star with the same central pressure is
\begin{equation}
\delta M_{grav}=m_{0}(R)+\frac{J^{2}}{R^{3}}
\end{equation}
The change in the radius of the star is given by
\begin{equation}
\delta R=\xi_{0}(R)+\xi_{2}(R)P_{2}(\theta)
\end{equation}
We wish to study sequences of the rotating stars with constant
total baryon number at variable spin frequency $\nu=\Omega/2\pi$.
The expansion of total baryon numbers in powers of $\Omega$ is
\begin{equation}
N_{B}=N_{B}^{0}+\delta N_{B}+O(\Omega^{4})
\end{equation}
where
\begin{equation}
N_{B}^{0}=\int_{0}^{R}n_{B}(r)[1-2M(r)/r]^{-1/2}4\pi r^{2}dr
\end{equation}
is the total baryons number of non-rotating star and
\begin{eqnarray}
\delta
N_{B}=\frac{1}{m_{N}}\int_{0}^{R}(1-\frac{2M(r)}{r})^{-1/2}\{[1+\frac{m_{0}(r)}
{r-2M(r)}+\frac{1}{3}r^{2}[\Omega-\omega(r)]^{2}e^{-\nu}]m_{N}n_{B}(r)\nonumber\\
+\frac{dm_{N}n_{B}(r)}{dP}(\epsilon+P) p_{0}^{*}(r)\}4\pi r^{2}dr
\end{eqnarray}
here $m_{N}$ is the rest mass per baryon. To construct constant
baryon number sequences, we first solve the TOV equations to find
the non-rotating configuration for giving a central pressure
P(r=0). And then, for an assigned value of the angular velocity
$\Omega$ the equations of star structure are solved to order
$\Omega^{2}$, imposing that the correction to the pressure $
p_{0}^{*}(r=0)$ being not equal to zero. The value of
$p_{0}^{*}(r=0)$ is then changed until the same baryon number as
that non-rotating star is obtained.

The results for the stability of rotating hybrid star
configurations with possible deconfinement phase transition
according to the EOS described above are shown in Fig.2, where the
total gravitation mass is given as functions of the equatorial
radius and the central baryon number density for static stars as
well as for stars rotating with the maximum rotation frequency
$\nu_{k}$. The dotted lines connect configuration with the same
total baryon number and it becomes apparent that the rotating
configurations are less compact than the static ones. In order to
explore the increase in central density due to spin down, we
create sequences of hybrid star models. Model in a particular
sequence have the same constant baryon number, increasing central
density and decreasing angular velocity. Fig.3 displays the
central density of rotating hybrid stars with different
gravitational mass at zero spin, as a function of its rotational
frequency. In the interior of these stars, the matter can be
gradually converted from the relatively incompressible nuclear
matter phase to more compressible quark matter phase.

\section*{3. Deconfinement heating}
As the star spins down, the centrifugal force decreases
continuously, increasing its internal density. At this occurrence,
the nuclear matter continuously converts into quark matter in an
exothermic reaction, i.e. $n\rightarrow u+2d, p\rightarrow 2u+d$,
s quarks immediately appear after weak decay. Fig.3 identifies the
fact that quarks are accumulating  in the interior of the star
with decreasing rotation frequency $\nu$. In a particular sequence
with constant baryon number, we can calculate the deconfinement
baryon number $N_{q}$ of the star with the varying rotation
frequency. The expression of $N_{q}$ is similar to that the total
baryon number $N_{B}$, except that the baryon number density
$n_{B}$ is displaced with the deconfinement baryon number density
$n_{q}$. Deconfinement baryon number of $1.4M_{\odot}$ star, for
example, is plotted in Fig.4. The analytic expression fits as
\begin{equation}
N_{q}=N_{q}^{0}(1-0.716\nu_{3}^{2}+0.055\nu_{3}^{3}-0.032\nu_{3}^{4})
\end{equation}
where $N_{q}^{0}\approx 0.22N_{\odot}$ is the baryon number of
quarks for the static configuration and $\nu_{3}=\nu/10^{3}$ Hz.
We can also derive similar expressions for sequences of hybrid
star models. Left panel in Fig.1 indicates release of latent heat
because there is the reduction in enthalpy per baryon at
transition. Assumed the average value of release energy per
nucleon that is transforming into quarks $q_{n}$, the total latent
heat per time can be written as
\begin{equation}
H_{dec}(t)=q_{n}\frac{dN_{q}}{d\nu}\dot{\nu}(t)
\end{equation}
with
\begin{equation}
\dot{\nu}=-\frac{16\pi^{2}}{3Ic^{3}}\mu^{2}\nu^{3}\sin^{2}\theta
\end{equation}
is induced by magnetic dipole radiation (MDR), where $I$ is the
stellar moment of inertia, $\mu=\frac{1}{2}BR^{3}$ is the magnetic
dipole moment, and $\theta$ is the inclination angle between
magnetic and rotational axes, $q_{n}$ can be estimated as the
order of 0.1 MeV through contrast of HP enthalpy to corresponding
MP enthalpy for same region of baryon number density, which is
uniquely defined for given stellar mass in our work.

\section*{4. Cooling curves}

The cooling is realized via two channels - by neutrino emission
from the entire star body and by transport of heat from the
internal layers to the surface resulting in the thermal emission
of photons. Neutrino emission is generated in numerous reactions
in the interiors of neutron stars, as reviewed, by Page et al.
\cite{18}. For the calculation of cooling of the hadron part of
the hybrid star we use the main processes which are nucleon direct
Urca (NDU) and nucleon modified Urca (NMU) and nucleon
bremsstrahlung (NB). If the proton and electron Fermi momenta are
too small compared with the neutron Fermi momenta, the NDU process
is forbidden because it is impossible to satisfy conservation of
momentum. Under typical conditions one finds that the ratio of the
number density of protons to that of nucleons must exceed about
0.11 for the process to be allowed. Medium effects and
interactions among the particles modify this value only slightly
but the presence of muons raise it to about 0.15. We apply the
second condition in our calculations. For the calculation of
cooling of the quark matter we consider the most efficient
processes: the quark direct Urca (QDU) processes on unpaired
quarks, the quark modified Urca (QMU) and the quark bremsstrahlung
(QB). Nucleon superfluidity and quark superconductivity are not
included in the model. Considering the energy equation of the
star, the cooling equation can be written as
\begin{equation}
C_{V}\frac{dT}{dt}=-L_{\nu}-L_{\gamma}+H
\end{equation}
where $C_{V}$ is the total stellar heat capacity, it incorporates
neutron contribution, quark contribution and electron
contribution. The term $H$ indicates the heating energy per unit
time, in our work $H=H_{dec}$, $L_{\nu}$ is the neutrino
luminosity, and $L_{\gamma}$ is the surface photon luminosity
given by
\begin{equation}
L_{\gamma}=4\pi R^{2}\sigma T_{s}^{4},
\end{equation}
here $\sigma$ is the Stefan-Boltzmann constant and $T_{s}$ is the
surface temperature. The surface temperature is related to
internal temperature by a coefficient determined by the scattering
processes occurring in the crust. We apply an formula which is
demonstrated by Gudmundsson et al. \cite{19},
\begin{equation}
T_{s}=3.08\times10^{6}g_{s,14}^{1/4}T_{9}^{0.5495}
\end{equation}
where $g_{s,14}$ is the proper surface gravity of the star in
units of $10^{14} cm s^{-2}$. The gravitational red-shift is also
taken into account. Then the effective surface temperature
detected by a distant observer is
$T_{s}^{\infty}=T_{s}\sqrt{1-R_{g}/R}$, where $R_{g}$ is the
gravitational stellar radius. In our calculation, we choose the
initial temperature $T_{0}=10^{9}$k, and the magnetic tilt angle
$\theta=45^{o}$.

 Fig.5 shows the cooling behavior of a 1.4$M_{\odot}$ hybrid star
 for different magnetic fields ($10^{9}-10^{13}$G).
In this figure, $q_{n}$ is taken to be 0.1 MeV. It is evident that
the DH increase the
 surface temperature dramatically. This is extremely different from
 fast cooling scenario (solid curve in Fig.5). We find that there
 is a quite clear magnetic-field dependence of the curves. It is
 determined by the properties of MDR. Solutions to equation (15)
 show the magnetic-field dependence of the spin frequencies. The
 strong field strength induces a rapid spin-down at the beginning
 while the low field strength leads to only obvious spin-down at
 the older ages (see Fig.3 in Zheng et al \cite{2}). Hence, a large
 quantity of nuclear matter is deconfined into quark matter during
 the earlier and shorter period for strong-field case. On the
 contrary, the plateau can form at late period for weak-field
 case.

 In Fig.6 we present the cooling behavior of different mass stars for B=$10^{12}$ with and without
 DH. The observational
 data, taken from tables 1 and 2 in Page et al.\cite{20}, have been
 shown in the figure. The theoretical curves are consistent with
 the observational data.
The heating rate has a regular mass
dependence. More massive neutron star have larger heating rate
 for the neutron stars gravitational mass $M \geq 1.2 M_{\odot}$.

 Fig.7 shows the evolution of the surface
 temperature of different mass stars for the weak (B=$10^{9}$G) magnetic
 field.
In the cases of weak field, stars could maintain high temperatures
even at older ages ($>10^{6}$yrs). This feature may give an
illustration of the inferred temperature for PSR J0437-4715
(Kargaltsev et al. \cite{21}), although the quantitative analysis
needs future study.

\section*{5. Conclusions and discussions}
The thermal evolution of rotating hybrid stars with DH have been
investigated in this work. Using Hartle's perturbative approach,
we have calculated the change of internal structure of rotating
hybrid stars. The nuclear matter can continuously be converted
into quark matter to release latent heat during the spins down of
star. The heat luminosity can be estimated as proportion to the
change rates of quark number. The results show the DH of rotating
stars leads to good agreement with the observed data in the case
of enhanced cooling (That is the onset of direct Urca process). We
also found that heating effect has magnetic field strength
dependence. For those stars with weak fields ($<10^{10}$G), our
results show that they can maintain a high temperature
($>10^{5}$K) at older ages ($t\sim10^{10}yrs$) such as PSR
J0437-4715. It may be particularly interesting for high
temperature of weak-field stars at late stage.

 We here calculate the released latent heat by
regarding the heat release per nucleon as a parameter. However,
the heat release per nucleon should be a density-dependent
quantity. Numerical calculations are necessary for heat
luminosity. Precise fittings need consider improvement of the star
model, the inclusions of such superfluidity, superconductivity and
tension effect in MP. These will be the future works.

\section*{Acknowledgments}
 This work is supported by NFSC under Grant Nos.90303007 and
10373007.

\clearpage

 \begin{figure}
   \centering
   \includegraphics[width=0.9\textwidth]{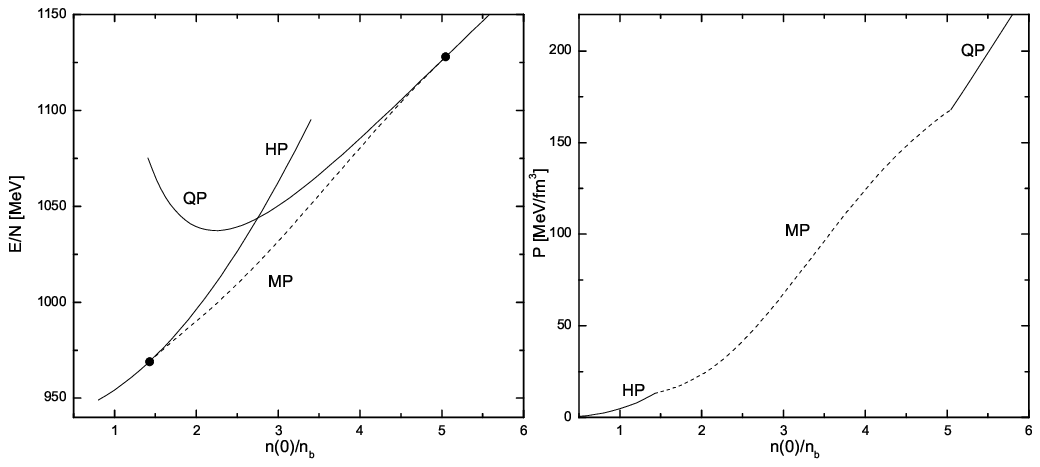}
   \caption{Model
EOS for energy per baryon and the pressure of hybrid star matter
as a function of the baryon number density. The HP EOS is a
relativistic mean-field model, the quark matter is
 effective mass MIT bag model with $m_{s}=150MeV$, $B^{1/4}=160 MeV$,coupling constant $g=3.0$.}

   \label{Fig:f1}
   \end{figure}

 \begin{figure}
   \centering
   \includegraphics[width=0.9\textwidth]{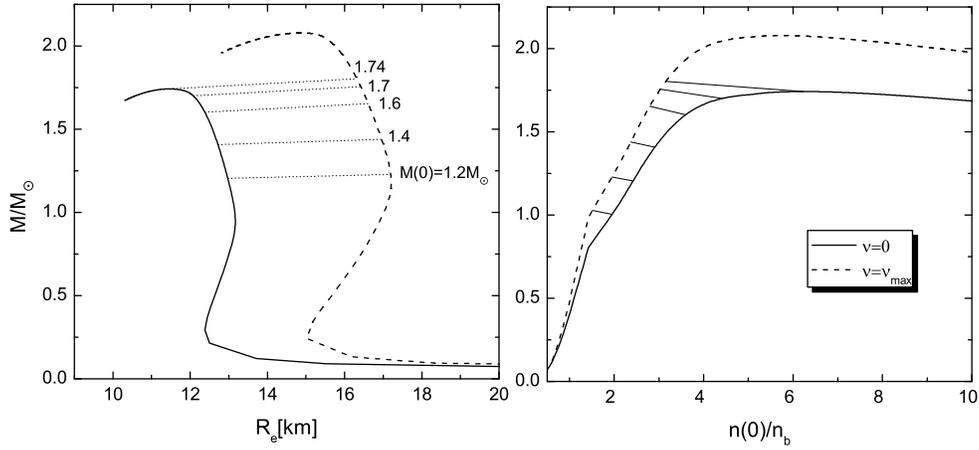}
   \caption{Gravitational mass M as a function of the equatorial radius (left figure) and the central
   density (right figure) for rotating hybrid stars configurations with a deconfinement phase phase transition.
 The solid curves correspond to static configurations. the dashed ones to those with maximum rotation frequency $\nu_{k}$.
 The lines between both extremal cases connect configurations with the same total baryon number.}
   \label{Fig:f2}
   \end{figure}

\begin{figure}
   \centering
   \includegraphics[width=0.6\textwidth]{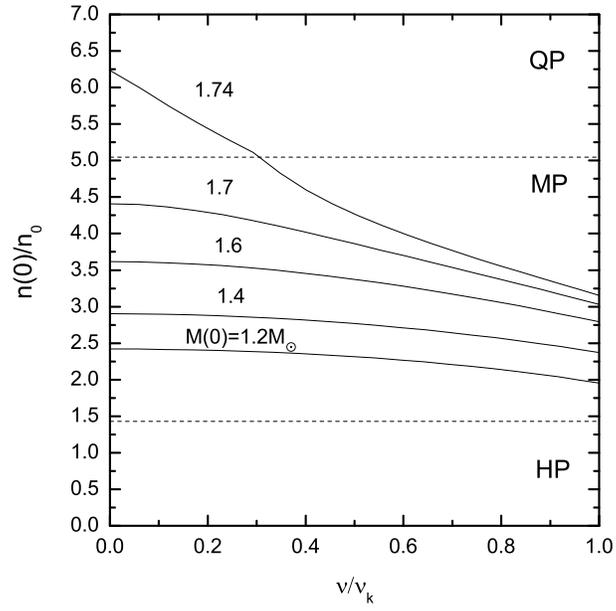}
   \caption{Central density as a function of rotational frequency for rotating hybrid
stars of different gravitational mass at zero spin. All sequences
are with constant total baryon number. Dash horizontal lines
indicate the density where quark matter is produced.
 }
   \label{Fig:f3}
   \end{figure}

\begin{figure}
\centering
   \includegraphics[width=0.6\textwidth]{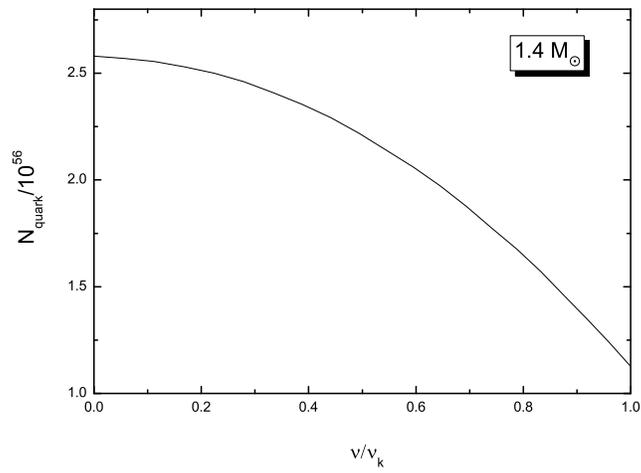}
   \caption{The number of converting quark into baryon as a function of rotational frequency for  1.4 $M_{\odot}$
  rotating hybrid star.}
   \label{Fig:f4}
   \end{figure}


\begin{figure}
   \centering
   \includegraphics[width=0.6\textwidth]{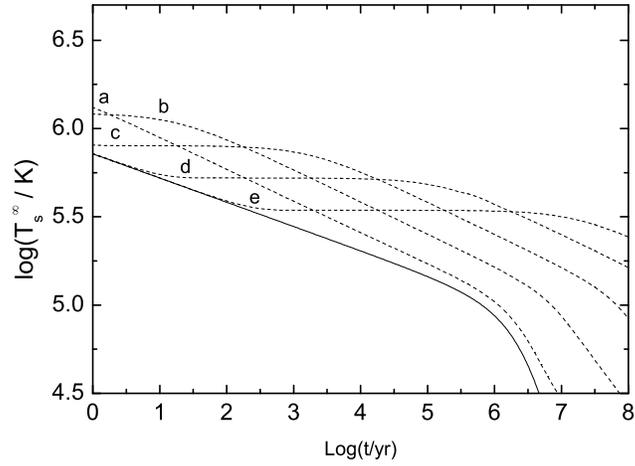}
   \caption{Cooling curves of 1.4 $M_{\odot}$ hybrid star with
   DH
   for various magnetic fields (curve a: $10^{13}$, b: $10^{12}$, c: $10^{11}$,
   d: $10^{10}$, e: $10^{9}$) and the curves without DH (solid
   curve).
 }
   \label{Fig:f6}
   \end{figure}

\begin{figure}
   \centering
   \includegraphics[width=0.6\textwidth]{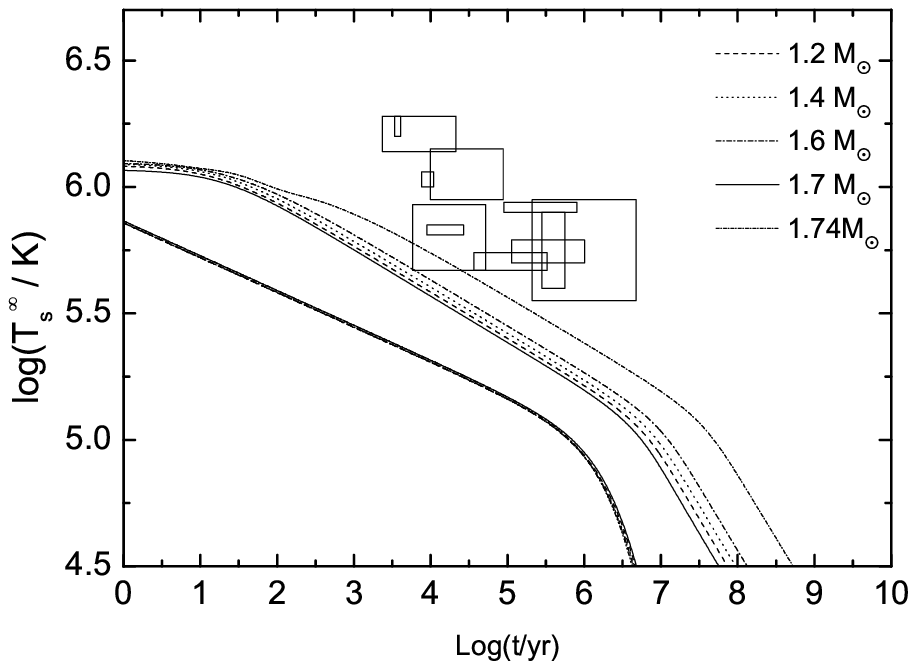}
   \caption{Cooling curves of neutron stars with DH for different star mass and B=$10^{12}$G (upper curves) and
   the curves without DH (lower curves).
 }
   \label{Fig:f6}
   \end{figure}
\begin{figure}
   \centering
   \includegraphics[width=0.6\textwidth]{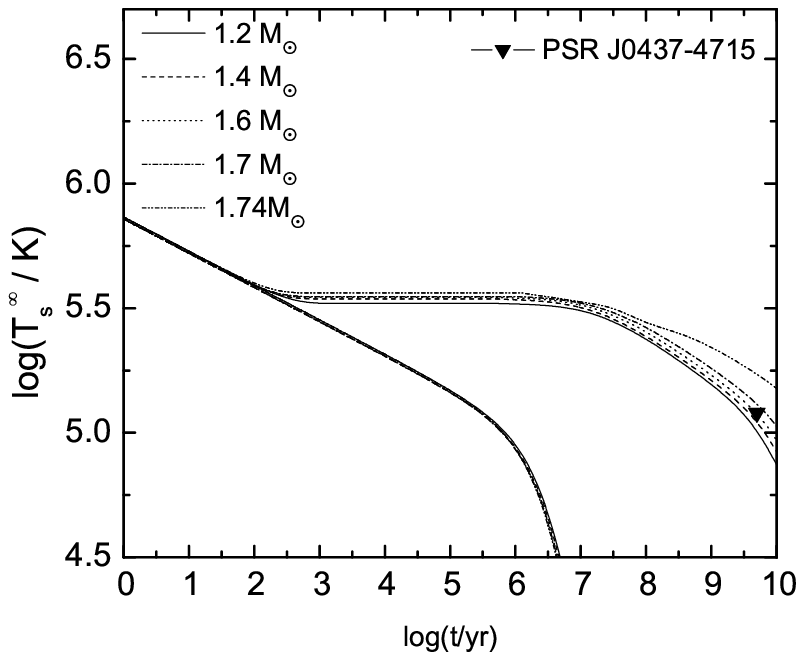}
   \caption{Cooling curves of neutron stars with DH for different star mass and B=$10^{9}$G (upper curves) and
   the curves without DH (lower curves).
 }
   \label{Fig:f6}
   \end{figure}


\begin{thebibliography}{}

\bibitem[\protect\citeauthoryear{{Abraham}, {Crawford} \& {McHardy}}
  {{Abraham}, {Crawford} \& {McHardy}}{1992}]
  {abraham+92}
  {Abraham} R.~G.,  {Crawford} C.~S.,    {McHardy} I.~M.,  1992, 
  \apj, 401, 474


\bibitem[\protect\citeauthoryear{{Bahcall}, {Kirhakos}, {Saxe} \& {Schneider}}
  {{Bahcall} et~al.}{1997}]
  {bahcall+97}
  {Bahcall} J.~N.,  {Kirhakos} S.,  {Saxe} D.~H.,    {Schneider} D.~P.,  1997,
  \apj, 479, 642

\bibitem[\protect\citeauthoryear{{Bernardi}, et~al.}
  {{Bernardi} et~al.}{2003}]
  {bernardi+03}
  {Bernardi} M., et~al., 
  2003, \aj, 125, 1849
  
\bibitem[\protect\citeauthoryear{{Block} \& {Stockton}}
  {{Block} \& {Stockton}}{1991}]
  {blockstockton91}
  {Block} D.~L.,  {Stockton} A.,  1991, 
  \aj, 102, 1928

\bibitem[\protect\citeauthoryear{{Blundell}, {Beasley}, {Lacy} \& {Garrington}}
  {{Blundell} et~al.}{1996}]
  {blundell+96}
  {Blundell} K.~M.,  {Beasley} A.~J.,  {Lacy} M.,    {Garrington} S.~T.,  1996,
  \apjl, 468, L91

\bibitem[\protect\citeauthoryear{{Blundell} \& {Rawlings}}{{Blundell} \&
    {Rawlings}}{2001}]{blundellrawlings01}
  {Blundell} K.~M.,  {Rawlings} S.,  2001, 
  \apjl, 562, L5

\bibitem[\protect\citeauthoryear{{Boyce}, et~al.}
  {{Boyce} et~al.}{1998}]
  {boyce+98}
  {Boyce} P.~J., et~al.,
  1998, \mnras, 298, 121

\bibitem[\protect\citeauthoryear{{Boyce}, {Disney} \& {Bleaken}}{{Boyce}
    et~al.}{1999}]{boyce+99}
  {Boyce} P.~J.,  {Disney} M.~J.,    {Bleaken} D.~G.,  1999, \mnras, 302, L39

\bibitem[\protect\citeauthoryear{{Dunlop}, {McLure}, {Kukula}, {Baum}, {O'Dea}
  \& {Hughes}}{{Dunlop} et~al.}{2003}]{dunlop+03}
{Dunlop} J.~S.,  {McLure} R.~J.,  {Kukula} M.~J.,  {Baum} S.~A.,  {O'Dea}
  C.~P.,    {Hughes} D.~H.,  2003, \mnras, 340, 1095

\bibitem[\protect\citeauthoryear{{Efstathiou}, {Ellis} \&
  {Peterson}}{{Efstathiou} et~al.}{1988}]{EEP88}
{Efstathiou} G.,  {Ellis} R.~S.,    {Peterson} B.~A.,  1988, \mnras, 232, 431

\bibitem[\protect\citeauthoryear{{Falomo}, {Kotilainen} \& {Treves}}{{Falomo}
  et~al.}{2001}]{falomo+01}
{Falomo} R.,  {Kotilainen} J.,    {Treves} A.,  2001, \apj, 547, 124

\bibitem[\protect\citeauthoryear{{Ferrarese} \& {Merritt}}{{Ferrarese} \&
  {Merritt}}{2000}]{ferraresemerritt00}
{Ferrarese} L.,  {Merritt} D.,  2000, \apjl, 539, L9

\bibitem[\protect\citeauthoryear{{Gebhardt}, et~al.}
  {{Gebhardt} et~al.}{2000}]
  {gebhardt+00}
  {Gebhardt} K.,  et~al.,
  2000, \apjl, 539, L13

\bibitem[\protect\citeauthoryear{{Goldschmidt}, {Kukula}, {Miller} \&
  {Dunlop}}{{Goldschmidt} et~al.}{1999}]{goldschmidt+99}
{Goldschmidt} P.,  {Kukula} M.~J.,  {Miller} L.,    {Dunlop} J.~S.,  1999,
  \apj, 511, 612

\bibitem[\protect\citeauthoryear{{Goldschmidt}, {Miller}, {La Franca} \&
  {Cristiani}}{{Goldschmidt} et~al.}{1992}]{goldschmidt+92}
{Goldschmidt} P.,  {Miller} L.,  {La Franca} F.,    {Cristiani} S.,  1992,
  \mnras, 256, 65P

\bibitem[\protect\citeauthoryear{{Green} \& {Yee}}{{Green} \&
  {Yee}}{1984}]{greenyee84}
{Green} R.~F.,  {Yee} H.~K.~C.,  1984, \apjs, 54, 495

\bibitem[\protect\citeauthoryear{{Gregory}, {Vavasour}, {Scott} \&
  {Condon}}{{Gregory} et~al.}{1994}]{gregory+94}
{Gregory} P.~C.,  {Vavasour} J.~D.,  {Scott} W.~K.,    {Condon} J.~J.,  1994,
  \apjs, 90, 173

\bibitem[\protect\citeauthoryear{{Hamilton}, {Casertano} \&
  {Turnshek}}{{Hamilton} et~al.}{2002}]{hamilton+02}
{Hamilton} T.~S.,  {Casertano} S.,    {Turnshek} D.~A.,  2002, \apj, 576, 61

\bibitem[\protect\citeauthoryear{{Hooper}, {Impey} \& {Foltz}}{{Hooper}
  et~al.}{1997}]{hooper+97}
{Hooper} E.~J.,  {Impey} C.~D.,    {Foltz} C.~B.,  1997, \apjl, 480, L95

\bibitem[\protect\citeauthoryear{{Hutchings}}{{Hutchings}}{1987}]{hutchings87}
{Hutchings} J.~B.,  1987, \apj, 320, 122

\bibitem[\protect\citeauthoryear{{Hutchings}, {Frenette}, {Hanisch}, {Mo},
  {Dumont}, {Redding} \& {Neff}}{{Hutchings} et~al.}{2002}]{hutchings+02}
{Hutchings} J.~B.,  {Frenette} D.,  {Hanisch} R.,  {Mo} J.,  {Dumont} P.~J.,
  {Redding} D.~C.,    {Neff} S.~G.,  2002, \aj, 123, 2936

\bibitem[\protect\citeauthoryear{{Hutchings}, {Johnson} \& {Pyke}}{{Hutchings}
  et~al.}{1988}]{hutchings+88}
{Hutchings} J.~B.,  {Johnson} I.,    {Pyke} R.,  1988, \apjs, 66, 361

\bibitem[\protect\citeauthoryear{{Hutchings} \& {Neff}}{{Hutchings} \&
  {Neff}}{1990}]{hutchings+90}
{Hutchings} J.~B.,  {Neff} S.~G.,  1990, \aj, 99, 1715

\bibitem[\protect\citeauthoryear{{Hutchings} \& {Neff}}{{Hutchings} \&
  {Neff}}{1991}]{hutchingsneff91}
{Hutchings} J.~B.,  {Neff} S.~G.,  1991, \aj, 101, 2001

\bibitem[\protect\citeauthoryear{{Hutchings} \& {Neff}}{{Hutchings} \&
  {Neff}}{1992}]{hutchingsneff92}
{Hutchings} J.~B.,  {Neff} S.~G.,  1992, \aj, 104, 1

\bibitem[\protect\citeauthoryear{{J{\o}rgensen}, {Franx} \&
  {Kjaergaard}}{{J{\o}rgensen} et~al.}{1996}]{jorgensen+96}
{J{\o}rgensen} I.,  {Franx} M.,    {Kjaergaard} P.,  1996, \mnras, 280, 167

\bibitem[\protect\citeauthoryear{{Kormendy} \& {Gebhardt}}{{Kormendy} \&
  {Gebhardt}}{2001}]{kormgeb01}
{Kormendy} J.,  {Gebhardt} K.,  2001, in 20th Texas Symposium on relativistic
  astrophysics {Supermassive Black Holes in Galactic Nuclei (Plenary Talk)}.

\bibitem[\protect\citeauthoryear{{Krist}}{{Krist}}{1999}]{tinytim}
{Krist} J.,  1999, TinyTim User Manual

\bibitem[\protect\citeauthoryear{{Kukula}, {Dunlop}, {McLure}, {Miller},
  {Percival}, {Baum} \& {O'Dea}}{{Kukula} et~al.}{2001}]{kukula+01}
{Kukula} M.~J.,  {Dunlop} J.~S.,  {McLure} R.~J.,  {Miller} L.,  {Percival}
  W.~J.,  {Baum} S.~A.,    {O'Dea} C.~P.,  2001, \mnras, 326, 1533

\bibitem[\protect\citeauthoryear{{Lehnert}, {van Breugel}, {Heckman} \&
  {Miley}}{{Lehnert} et~al.}{1999}]{lehnert+99b}
{Lehnert} M.~D.,  {van Breugel} W.~J.~M.,  {Heckman} T.~M.,    {Miley} G.~K.,
  1999, \apjs, 124, 11

\bibitem[\protect\citeauthoryear{{M{\' a}rquez}, {Petitjean}, {Th{\' e}odore},
  {Bremer}, {Monnet} \& {Beuzit}}{{M{\' a}rquez} et~al.}{2001}]{marquez+01}
{M{\' a}rquez} I.,  {Petitjean} P.,  {Th{\' e}odore} B.,  {Bremer} M.,
  {Monnet} G.,    {Beuzit} J.-L.,  2001, \aap, 371, 97

\bibitem[\protect\citeauthoryear{{Magorrian}, et~al.}
  {{Magorrian} et~al.}{1998}]
  {magorrian+98}
  {Magorrian} J., et~al., 
  1998, \aj, 115, 2285

\bibitem[\protect\citeauthoryear{{Malkan}}{{Malkan}}{1984}]{malkan84}
{Malkan} M.~A.,  1984, \apj, 287, 555

\bibitem[\protect\citeauthoryear{{Marconi}, {Axon}, {Macchetto}, {Capetti},
  {Soarks} \& {Crane}}{{Marconi} et~al.}{1997}]{marconi+97}
{Marconi} A.,  {Axon} D.~J.,  {Macchetto} F.~D.,  {Capetti} A.,  {Soarks}
  W.~B.,    {Crane} P.,  1997, \mnras, 289, L21

\bibitem[\protect\citeauthoryear{{Marconi} \& {Hunt}}{{Marconi} \&
  {Hunt}}{2003}]{marconihunt03}
{Marconi} A.,  {Hunt} L.~K.,  2003, \apjl, 589, L21

\bibitem[\protect\citeauthoryear{{McLeod} \& {Rieke}}{{McLeod} \&
    {Rieke}}{1995}]{mcleodrieke95} 
{McLeod} B.~A.,  {Rieke} G.~H.,  1995, \apj, 441, 96

\bibitem[\protect\citeauthoryear{{McLeod} \& {McLeod}}{{McLeod} \&
  {McLeod}}{2001}]{mcleod01}
{McLeod} K.~K.,  {McLeod} B.~A.,  2001, \apj, 546, 782

\bibitem[\protect\citeauthoryear{{McLeod}, {Rieke} \&
  {Storrie-Lombardi}}{{McLeod} et~al.}{1999}]{mcleod+99}
{McLeod} K.~K.,  {Rieke} G.~H.,    {Storrie-Lombardi} L.~J.,  1999, \apjl, 511,
  L67

\bibitem[\protect\citeauthoryear{{McLure} \& {Dunlop}}{{McLure} \&
  {Dunlop}}{2002}]{mcluredunlop02}
{McLure} R.~J.,  {Dunlop} J.~S.,  2002, \mnras, 331, 795

\bibitem[\protect\citeauthoryear{{McLure}, {Dunlop} \& {Kukula}}
  {{McLure} et~al.}{2000}]
  {mclure+00}
  {McLure} R.~J.,  {Dunlop} J.~S.,    {Kukula} M.~J.,  2000, \mnras, 318, 693

\bibitem[\protect\citeauthoryear{{McLure}, {Kukula}, {Dunlop}, {Baum}, {O'Dea}
  \& {Hughes}}{{McLure} et~al.}{1999}]{mclure+99}
{McLure} R.~J.,  {Kukula} M.~J.,  {Dunlop} J.~S.,  {Baum} S.~A.,  {O'Dea}
  C.~P.,    {Hughes} D.~H.,  1999, \mnras, 308, 377

\bibitem[\protect\citeauthoryear{{Percival}, {Miller}, {McLure} \&
  {Dunlop}}{{Percival} et~al.}{2001}]{percival+01}
{Percival} W.~J.,  {Miller} L.,  {McLure} R.~J.,    {Dunlop} J.~S.,  2001,
  \mnras, 322, 843

\bibitem[\protect\citeauthoryear{{Press}, {Teukolsky}, {Vetterling} \&
  {Flannery}}{{Press} et~al.}{1992}]{numrec}
{Press} W.~H.,  {Teukolsky} S.~A.,  {Vetterling} W.~T.,    {Flannery} B.~P.,
  1992, Numerical recipes in FORTRAN. The art of scientific computing.
Cambridge: University Press, 1992, 2nd ed.

\bibitem[\protect\citeauthoryear{{Puchnarewicz}, et~al.}
  {{Puchnarewicz} et~al.}{1992}]
  {puchnarewicz+92}
  {Puchnarewicz} E.~M.,  et~al., 
  1992, \mnras, 256, 589

\bibitem[\protect\citeauthoryear{{Reimers}, et~al.}
  {{Reimers} et~al.}{1995}]
  {reimers+95}
  {Reimers} D.,  et~al.,
  1995, \aap, 303, 449

\bibitem[\protect\citeauthoryear{{Ridgway}, {Heckman}, {Calzetti} \&
  {Lehnert}}{{Ridgway} et~al.}{2001}]{ridgway+01}
{Ridgway} S.~E.,  {Heckman} T.~M.,  {Calzetti} D.,    {Lehnert} M.,  2001,
  \apj, 550, 122

\bibitem[\protect\citeauthoryear{{S\'{e}rsic}}{{S\'{e}rsic}}{1968}]{sersic68}
{S\'{e}rsic} J.~L.,  1968, in Atlas de Galaxes Australes; Vol. Book; Page 1 {Atlas de Galaxes Australes}.

\bibitem[\protect\citeauthoryear{{Smith}, {Heckman}, {Bothun}, {Romanishin} \&
  {Balick}}{{Smith} et~al.}{1986}]{smith+86}
{Smith} E.~P.,  {Heckman} T.~M.,  {Bothun} G.~D.,  {Romanishin} W.,    {Balick}
  B.,  1986, \apj, 306, 64

\bibitem[\protect\citeauthoryear{{Stockton} \& {Ridgway}}{{Stockton} \&
  {Ridgway}}{2001}]{stocktonridgway01}
{Stockton} A.,  {Ridgway} S.~E.,  2001, \apj, 554, 1012

\bibitem[\protect\citeauthoryear{{Tadhunter}, {Marconi}, {Axon}, K., {Robinson}
  \& {Jackson}}{{Tadhunter} et~al.}{2003}]{tadhunter+03}
{Tadhunter} C.,  {Marconi} A.,  {Axon} D.,  K. W.,  {Robinson} T.~G.,
  {Jackson} N.,  2003, \mnras

\bibitem[\protect\citeauthoryear{{Veron-Cetty} \& {Woltjer}}{{Veron-Cetty} \&
  {Woltjer}}{1990}]{veron90}
{Veron-Cetty} M.~.,  {Woltjer} L.,  1990, \aap, 236, 69

\bibitem[\protect\citeauthoryear{{Veron-Cetty} \& {Veron}}{{Veron-Cetty} \&
  {Veron}}{1993}]{VCV1993}
{Veron-Cetty} M.-P.,  {Veron} P.,  1993, {A Catalogue of quasars and active
  nuclei}.
ESO Scientific Report, Garching: European Southern Observatory (ESO), |c1993,
  6th ed.

\bibitem[\protect\citeauthoryear{{Veron-Cetty} \& {Veron}}{{Veron-Cetty} \&
  {Veron}}{2000}]{VCV2000}
{Veron-Cetty} M.-P.,  {Veron} P.,  2000, {A catalogue of quasars and active
  nuclei}.
A catalogue of quasars and active nuclei / M.-P.~Veron-Cetty and P.~Veron.~
  Garching bei Munchen, Germany : European Southern Observatory,
  c2000.~(Scientific report (European Southern Observatory) ; no.~19)

\bibitem[\protect\citeauthoryear{{Voges}, {Aschenbach}, {Boller}, {Br{\"
  a}uninger}, {Briel} \& {Burkert}}{{Voges} et~al.}{1999}]{voges+99}
{Voges} W.,  {Aschenbach} B.,  {Boller} T.,  {Br{\" a}uninger} H.,  {Briel} U.,
     {Burkert} W.,  1999, \aap, 349, 389

\bibitem[\protect\citeauthoryear{{Wright}, {McHardy} \& {Abraham}}{{Wright}
  et~al.}{1998}]{wright+98}
{Wright} S.~C.,  {McHardy} I.~M.,    {Abraham} R.~G.,  1998, \mnras, 295, 799

\bibitem[\protect\citeauthoryear{{Wyckoff}, {Gehren} \& {Wehinger}}{{Wyckoff}
  et~al.}{1981}]{wyckoff+81}
{Wyckoff} S.,  {Gehren} T.,    {Wehinger} P.~A.,  1981, \apj, 247, 750

\bibitem[\protect\citeauthoryear{{Yee} \& {Green}}{{Yee} \&
  {Green}}{1987}]{yeegreen87}
{Yee} H.~K.~C.,  {Green} R.~F.,  1987, \apj, 319, 28

\end{thebibliography}


\begin{thebibliography}{}
\bibitem[1992]{4} Cheng, K. S., Chau, W. Y., Zhang, J. L., \& Chau, H. F. 1992, ApJ, 396, 235
\bibitem[2000]{17}Chubarian E., Grigorian H., Poghosyan G., Blaschke D., 2000, A\&A, 357, 968
\bibitem[1987]{14}Gavai R. V., Potvin J., Sanielevici S., 1987
Phys.Rev.Lett, 58, 2519
 \bibitem[1997]{11} Glendenning. N.K. Compact Stars(Springer-verlag). 1997
 \bibitem[1992]{15}Glendenning N. K., 1992, Phys. Rev. D, 46, 1274
 \bibitem[1983]{19} Gudmundsson, E. H., Pethick, C. J.,\&
 Epstein, R. I. 1983, ApJ, 272,286.
\bibitem[1991]{7} Haensel P., Zdunik J., 1991, In: Madsen J.,
Haensel P. (eds.) Strange Quark Matter in Physics and
Astrophysics. (Nucl. Phys. B [Proc.Suppl.] 24),139
 \bibitem[1967]{10}Hartle J. B., 1967, ApJ, 150, 1005
\bibitem[2004]{21} Kargaltsev, O., Pavlov, G. G., \& Romani. R.
2004, APJ, 602,327
\bibitem[2005]{18} Page, D., Geppert, U., Weber, F. 2005,
 astro-ph/0508056.
\bibitem[2004]{20} Page, D., Lattimer,J. M., Prakash, M.,\&
Steiner, A. W. 2004, ApJS, 155, 623.
\bibitem[1984]{13}Pisalski R. D., Wilczek F., 1984, Phys.
Rev. Lett, 29, 338
 \bibitem[1995]{5} Reisenegger, A. 1995, ApJ, 442, 749
\bibitem[2006]{6} Reisenegger, A. Jofre, P., Fernandez, R., Kantor, E. 2006, APJ(in press)
[astro-ph/0606322]
\bibitem[1997]{12} Schertler, K., Greiner, C., Thoma, M.H., 1997,
Nucl. Phys. A616,659
 \bibitem[1989]{3} Shibazaki, N., \& Lamb, F. K. 1989, ApJ, 346, 808
\bibitem[1999]{8} Yuan, Y. F., Zhang, J. L. 1999, A\& A,
 344,371

\bibitem[2006]{9} Yu, Y. W., \& Zheng, X. P. 2006, A\&A, 445,
627
\bibitem[2006]{2} Zheng, X. P., Yu, Y. W., 2006, MNRAS, 369,376


\end{thebibliography}
\end{document}